# Electron Beam-Induced Nanopores in Bernal-Stacked Hexagonal Boron Nitride

Mehmet Dogan[1,2 (*)], S. Matt Gilbert[1,2,3 (*)], Thang Pham[1,2,3,4], Brian Shevitski[1,2,3,5], Peter Ercius[5], Shaul Aloni[5], Alex Zettl[1,2,3,#], Marvin L. Cohen[1,2,#]

[1] Department of Physics, University of California, Berkeley, CA 94720, USA
[2] Materials Sciences Division, Lawrence Berkeley National Laboratory, Berkeley, CA 94720, USA
[3] Kavli Energy NanoScience Institute at the University of California, Berkeley, and the Lawrence Berkeley National Laboratory, Berkeley, CA 94720, USA
[4] Department of Materials Science and Engineering, University of California, Berkeley, CA 94720, USA
[5] Molecular Foundry, Lawrence Berkeley National Laboratory, Berkeley, CA 94720, USA
[(*)] These authors contributed equally to this work.
[(#)] To whom correspondence should be addressed: azettl@berkeley.edu, mlcohen@berkeley.edu

**Abstract**

Controlling the size and shape of nanopores in two-dimensional materials is a key challenge in applications such as DNA sequencing, sieving, and quantum emission in artificial atoms. We here investigate experimentally and theoretically triangular vacancies in (unconventional) Bernal-stacked AB-*h*-BN formed using a high-energy electron beam. Due to the geometric configuration of AB-*h*-BN, triangular pores in different layers are aligned, and their sizes are controlled by the duration of the electron irradiation. Interlayer covalent bonding at the vacancy edge is not favored, as opposed to what occurs in the more common AA′-stacked BN. A variety of monolayer, concentric and bilayer pores in bilayer AB-*h*-BN are observed in high-resolution transmission electron microscopy and characterized using *ab initio* simulations. Bilayer pores in AB-*h*-BN are commonly formed, and grow without breaking the bilayer character. Nanopores in AB-*h*-BN exhibit a wide range of electronic properties, ranging from half-metallic to non-magnetic and magnetic semiconducting. Therefore, because of the controllability of the pore size, the electronic structure is also highly controllable in these systems, and can potentially be tuned for particular applications.

Vacancy defects in two-dimensional materials are zero-dimensional features that can impart to the host material optical and electronic properties (e.g. bandgaps, charge state, and electron scattering behavior) very different from those of the pristine sheets, and they can change the emergent measurable properties of the material (e.g. resistivity and tensile strength).[1–5] Moreover, such vacancies represent a physical structure in which there may be a small hole (i.e. nanopore) in an otherwise impermeable membrane.

Advances in atomic-resolution imaging and large-scale synthesis of two-dimensional materials have resulted in a rapid increase in the understanding of vacancy defects in these materials.[6–8] These vacancies have been studied in both naturally occurring and artificially produced forms.[2,9–14] The fundamental properties of vacancy defects in two-dimensional materials have led to their investigation in numerous research directions, including DNA sequencing,[15–20] quantum emission,[21] and molecular sieving.[22–27]

Vacancy size and geometry are critical for determining the utility of nanopores. While high-energy electron beams and ion beams have been used to create the smallest pores observed, the shape and size of nanopores is limited by the beam geometry.[14,18,28,29] In graphene, for example, pore size has traditionally been limited to 3 nm or greater and their shapes can only be minimally controlled.[18] This has been overcome in *h*-BN by leveraging the intrinsic etching properties of the material. When *h*-BN is exposed to 80 kV electron irradiation in TEM, vacancies are etched, and atomically precise edges are formed along the nitrogen zig-zag direction,[6,10,30–37] which is generally thought to be caused by the difference in the electron knock-on energy thresholds of boron atoms (74 kV) and nitrogen atoms (84 kV).[10,11,30,38,39]

The computed atomic structure of a boron monovacancy in monolayer *h*-BN is shown in **Figure 1a**. After its formation, under continuing electron irradiation, one of the under-coordinated nitrogen atoms adjacent to the vacancy gets ejected, followed by its neighboring boron atoms, which results in a triangular tetravacancy with a nitrogen edge. Maintaining the electron irradiation further results in progressively larger triangular vacancies.[35,37,39,40] This mechanism provides a robust method of creating nanopores that are equilateral triangles with the desired size by controlling the duration of the irradiation. The size, in turn, determines the electronic structure of these triangular pores, resulting in metallic, half-metallic or semiconducting edge regions.[41–47]

However, the ability to control the nanopore shape and size does not extend to samples thicker than a monolayer in conventional AA′-stacked *h*-BN (AA′-*h*-BN). As we will demonstrate, Bernal-stacked *h*-BN (AB-*h*-BN) does not have this impediment, enabling us to create triangular pores, both in a single layer of a thicker sample, and in multiple layers with concentric or fully aligned pore edges. We will describe how this is achieved, and characterize the electronic properties of these pores using first-principles calculations.

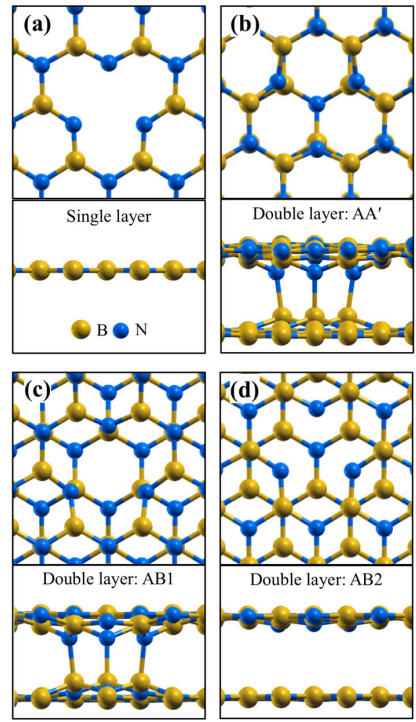

**Figure 1.** Relaxed atomic structure of a boron monovacancy in (a) monolayer, (b) bilayer AA′-, (c) bilayer AB1- and (d) bilayer AB2-*h*-BN. For each structure, top and side views of the atomic configuration are presented. The calculations are conducted in electron-rich conditions (3 extra electrons per 5×5 cell).

In multilayer *h*-BN, vacancy defects can interact with adjacent layers,[3,4] which can affect the types and shapes of vacancies in the sheets. In AA′-*h*-BN, interlayer interactions can dominate the properties of its electron-beam-induced defects. Out-of-plane covalent bonds form across layers at boron monovacancy sites in bilayers.[3] Also, along the edges of larger vacancies, covalent bonds form between the layers and the edge relaxes to resemble an *h*-BN nanotube with a small radius of curvature.[4]

Each lattice site in AA′-*h*-BN contains a boron directly under a nitrogen, thus interlayer covalent bonds are available at each site in the crystal. When a B monovacancy is formed, some or all of the three under-coordinated N atoms neighboring the vacancy may form bonds with the B atoms directly underneath (**Figure 1b**). We find that in the electron-rich regime (3 extra electrons per 5×5 cell as in Reference 3), forming these interlayer bonds is energetically favorable, with 2 out of 3 bonds being the most favorable configuration (**Table S1**). Therefore, the shape of the vacancy is not fully determined by its size, and the three-fold rotational symmetry of the lattice may be lost, which has been observed in AA′-*h*-BN.[3] When all 3 of the interlayer bonds are formed, the vacancy becomes a semiconductor, losing the half-metallic character of the B monovacancy edge (**Figure S1**). Furthermore, in the AA′-*h*-BN, the second layer is obtained from the first layer by a 60° rotation, thus the triangular pores with N edge in successive layers are not aligned, and multilayer pores do not have a pre-determined shape (**Figure S2**). This is consistent with the observations in References 6,10,30.

In Reference 40, we described a synthesis technique for reliably producing AB-*h*-BN. For this stacking sequence only half of the atoms are part of boron-nitrogen stacks. In **Figure 1c,d**, we present the structure of a boron monovacancy in one layer of double layer AB-*h*-BN. As opposed to the AA′ stacking, in AB-*h*-BN the two layers are inequivalent: in **Figure 1c**, the B atoms in the top layer are aligned with the hollow sites in the bottom layer, whereas the B atoms in the bottom layer are aligned with the N atoms in the top layer. Thus, a vacancy in the top layer yields a different configuration from a vacancy in the bottom layer. In order to consider vacancies only in the top layer, we can flip the system over if the vacancy is in the bottom layer, and give the resulting stacking a different name (AB1 for **Figure 1c** and AB2 for **Figure 1d**). The two stacking sequences are physically equivalent, and any vacancy configuration in one can be obtained in the other by rotating the system by 180° around an armchair direction. Distinguishing AB1 and AB2 allows us to meaningfully distinguish a top layer and a bottom layer in the discussions that follow.

In AB1-*h*-BN, a B monovacancy in the top layer creates three under-coordinated N atoms that are directly on top of B atoms from the bottom layer, which yields interlayer bonding in the electron-rich regime, as in the AA′ case (**Figure 1c** and **Table S1**). However, in AB2-*h*-BN, the under-coordinated N atoms lack a neighbor directly underneath, and thus the top layer stays flat and no new bonds form (see **Figure 1d**). Decoupling the two layers retains the magnetic properties of each layer, the half-metallic nature of the B monovacancy in the top layer (**Figure S1**).

In general, when the two layers remain chemically decoupled, their single-layer electronic and magnetic properties are retained, which can allow for the size-dependent magnetism found in single-layer *h*-BN nanopores to exist within a multilayer structure without disturbance. In **Figure 2.**, we summarize the dependence of the magnetic properties on the pore size based on our calculations. For the B monovacancy (pore size: 1) (**Figure 2. a**), each of the three neighboring N atoms is magnetized. For the pore that consists of 3 B and 1 N vacancy (tetravacancy, pore size: 2) (**Figure 2. b**), magnetism is zero due to the dimerization of the corner N atoms. For larger pores (pore size: 3, 4, 5,…), all the non-corner edge N atoms are magnetized and the corners remain dimerized (**Figure 2. c,d**). Each magnetized N atom carries a magnetization of approximately 1 $\mu_B$ (the exact values for the edge N are printed in **Figure 2.**, and see **Table S2** and **Table S3** for the remaining values carried by the neighboring atoms in certain cases). Therefore, the total magnetization (in $\mu_B$) of a pore is $M_{\text{total}} \cong$ 3, 0, 3, 6, 9, 12, ... for pore sizes 1, 2, 3, and so on.

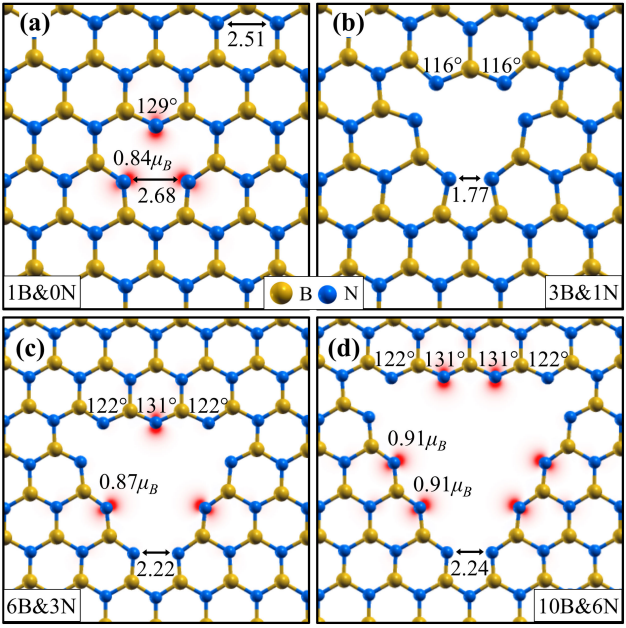

**Figure 2.** N-terminated triangular nanopores in single-layer *h*-BN with (a) pore size: 1, (b) pore size: 2, (c) pore size: 3 and (d) pore size: 4. For each chase, spatial distribution of computed magnetization is also plotted. For significantly magnetized atoms, magnetization values (computed by decomposing the Löwdin charges on the atomic orbitals) are printed next to representative atoms (there is three-fold symmetry in each case). Some important angles and distances (in Å) are also shown.

To realize these pores experimentally in multilayer *h*-BN, we grow AB-*h*-BN using chemical vapor deposition (CVD) on Cu and Fe substrates, and fabricate nanopores using an 80 kV electron beam (see Supplementary Material for further details). Studying our AB-*h*-BN samples using high-resolution transmission electron microscopy (HRTEM) focal series reconstruction,[3,6,40] we find many monolayer vacancies in two-layer stacks. **Figure 3** shows three such monolayer vacancies in the phase of the reconstructed exit wave. Assuming that the pores are formed in the top layer, the first one corresponds to AB2 stacking with a boron monovacancy (**Figure 3a**). In the second one, a tetravacancy is formed in the top layer of an AB1 bilayer (**Figure 3b**). The third one corresponds to a larger nanopore with 10 B and 6 N atoms missing from the top layer in bilayer AB2-*h*-BN (**Figure 3c**). The relaxed theoretical structures (which contain no interlayer covalent bonds) match the experimental HRTEM structure almost exactly, as shown in the middle row (simulated HRTEM exit waves based on the DFT-computed coordinates are presented in **Figure S3**). Spin-resolved DFT calculations result in different characterizations for these vacancies: half-metal with a total magnetization of 3.26 $\mu_B$ (**Figure 3a**); non-magnetic semiconductor with two sharp in-gap states 1.8 eV and 2.2 eV above the valence band maximum (**Figure 3b**); and magnetic semiconductor with gaps of 0.6 eV and 0.2 eV for the two spin channels and a total magnetization of 6.00 $\mu_B$ (**Figure 3c**).

This result suggests that tuning vacancy size in AB-*h*-BN is a reliable method for tuning its magnetic and electronic properties. The ability to create vacancies with tunable bandgaps in the visible range may be useful for h-BN in optoelectronics and photon emission applications.[21,48]

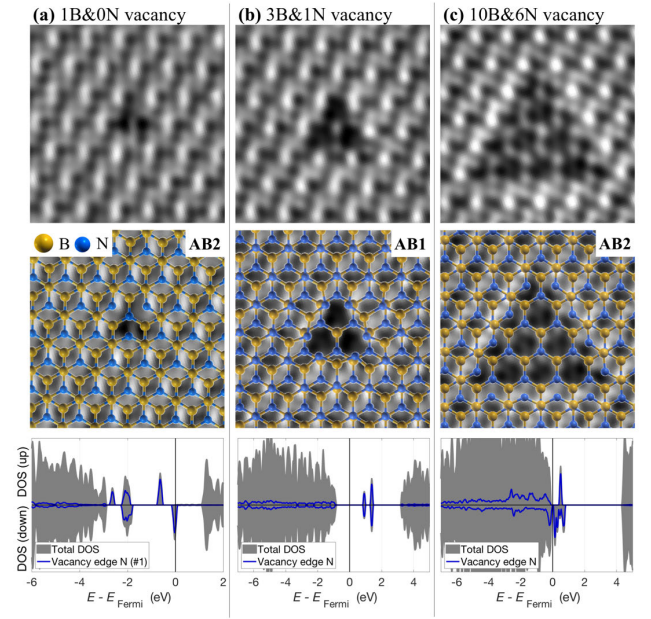

**Figure 3.** Top row: Phase of HRTEM focal series reconstructions of vacancies produced in a single layer of a bilayer AB-stacked h-BN, comprised of

(a) 1 B atom, (b) 3 B & 1 N atoms, and (c) 10 B & 6 N atoms. Each field-of-view is 2 nm wide. Middle row: Same images with the computed atomic structures overlaid. Bottom row: Computed spin-resolved density of states (DOS) plots. The total DOS is plotted as gray areas, and the blue curves are obtained by adding up the DOS projected onto the 2s and 2p orbitals of the under-coordinated N atoms at the vacancy edges.

In a multilayer AB-*h*-BN sample, because there is no relative rotation between layers,[40] the triangular pores with N edge are parallel regardless of which layer they are in, which is not the case for AA′-*h*-BN (**Figure S2**). After a monolayer triangular vacancy is formed, if the electron irradiation is maintained, a triangular vacancy tends to form in the exposed region of the other layer, resulting in nested triangular vacancies. **Figure S4** shows the growth of a vacancy in a monolayer of *h*-BN as it merges with a larger vacancy in the next layer to form a bilayer nanopore: When a small one- or few-atom vacancy forms within a triangular monolayer region that is embedded within a bilayer area, first, two of its edges align with the large vacancy to form bilayer edges, then the third edge aligns to form a "bilayer pore", and then the bilayer pore grows while retaining its bilayer edges and triangular geometry.

A possible mechanism that leads to this inclination to form bilayer edges is that a covalent bond forms between the edge atoms. In this scenario, as the triangular vacancies in the top and bottom *h*-BN layers intersect, the edge atoms covalently bond together. This would form a stable edge that maintains its structure as it is irradiated and reforms as it is etched. Below we take a closer look at bilayer edges to examine this hypothesis as well as alternate mechanisms.

We have performed HRTEM on bilayer triangular pores and edges in bilayer AB-*h*-BN, and found that when we are able to identify atoms and stacks of atoms in a bilayer edge image, we systematically obtain the same edge configuration. In **Figure 4**, we present a representative image for this configuration. **Figure 4a** shows an overview of two 3 nm bilayer vacancies in the AB stacked *h*-BN as highlighted by the green triangles. Multiple smaller monolayer vacancies are also present. The edge highlighted by the blue dashed box is presented in **Figure 4b**. In order to investigate any potential lattice relaxation in the structures, we overlay a grid that has spacings $a_0/2$ in the *x*-direction and $\sqrt{3}a_0/4$ in the *y*-direction, where $a_0$ is the lattice constant. All of the atomic positions for the undisturbed lattice fall on this grid, as shown in **Figure 4c**. We observe that double-atom stacks of boron and nitrogen (brightest peaks in the image) fall exactly on the grid of the undisturbed lattice, independent of how close they are to the edge (**Figure 4b**). This indicates that interlayer relaxation and covalent bond formation is unlikely in these bilayer edges. We note that we are not always able to identify atoms and stacks of atoms in the HRTEM images of bilayer edges, and thus we do not entirely rule out other configurations.

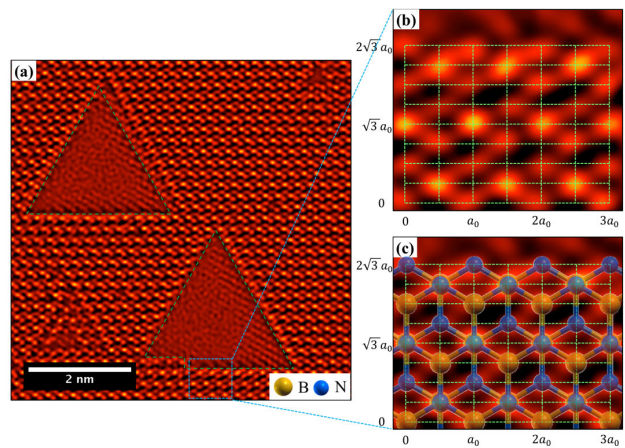

**Figure 4.** (a) HRTEM focal series reconstruction image of triangular vacancies produced in h-BN under 80 kV electron irradiation. Two ~3-nm bilayer pores (highlighted in green) are present with several smaller monolayer vacancies. The image is presented using a high-contrast filter to distinguish between vacuum (black and dim red), single boron atoms (bright red dots), and stacks of boron and nitrogen (brighter yellowish dots). (b) A zoomed-in section of the edge highlighted by blue dashed box in (a). An overlaid grid denotes the atomic positions for the undisturbed bilayer *h*-BN lattice. (c) The same zoomed-in section of the edge from (a) with the atomic species in an undisturbed lattice overlaid.

To further understand these bilayer edges, we have computed possible edge structures. The best candidate for the bilayer edge in **Figure 4** is presented in **Figure 5**. In this "open edge" case, two monolayer edges sit on top of one another without interlayer bonding. Other types of bilayer edges may also be present in these systems, and our calculations regarding them, including some "closed edge" cases, will be presented in a future report. The unrelaxed grid of positions is overlaid on the open edge structure as in **Figure 4**, showing that in-plane distortions are negligible. In previous computational studies, the nitrogen-terminated zigzag edge of the BN monolayer was found to be half-metallic.[43,49] Looking at the projected density of states (PDOS) plots in **Figure 5**, we observe that the open bilayer edge preserves the half-metallic character of the monolayer edge, with a magnetization of 2.02 $\mu_B$ per cell along the edge direction, localized on the edge N atoms (localization of the magnetization is depicted in **Figure S5**).

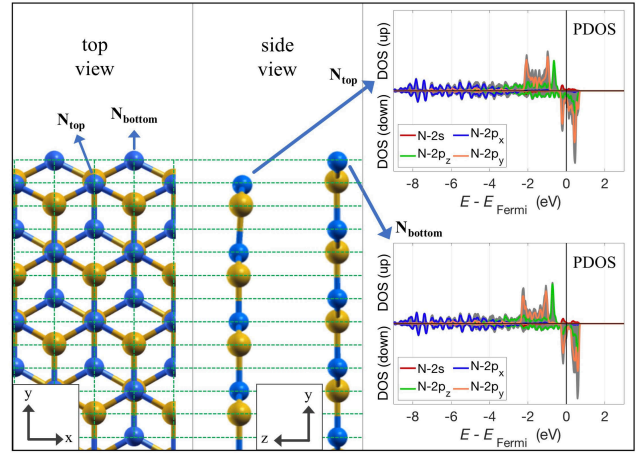

**Figure 5.** The nitrogen-terminated zigzag "open edge" and its projected density of states (PDOS) for the edge atoms. For the atomic models, an overlaid grid denotes the atomic positions for the undisturbed bilayer *h*-BN lattice.

Since closed edges are not regularly observed in our images, they are not the explanation of why the pores preferentially form a bilayer edge. We suggest that the explanation may be that bare monolayers of *h*-BN are less stable under electron irradiation. A second layer may provide protection from chemical sputtering from the electron beam or may increase the kinetic scattering threshold for the layers. Thus, whenever a small monolayer region of *h*-BN is exposed, the layer is etched back to form a bilayer edge. Therefore, the formation of bilayer edges in AB-*h*-BN appears to be due to kinetic effects, whereas, in AA′-*h*-BN, bilayer edges occur because of stabilizing interlayer bonds. As a result, bilayer edges in AB-*h*-BN greatly differ from the bilayer edges in AA′-stacked BN, which are insulating and have large in-plane and out-of-plane relaxations.[4] This is an important difference as it suggests that AB-*h*-BN could potentially be used for its 1D conducting edges.

We have achieved the controlled formation of triangular vacancies in multilayer *h*-BN, using a growth technique which results in Bernal stacking (AB-*h*-BN). Due to the favorability of nitrogen terminated vacancy edges in BN and the lack of relative rotation between the layers in Bernal stacking, triangular pores in different layers are aligned in AB-*h*-BN. We have shown that interlayer covalent bonding following vacancy formation in a layer is not favored for this stacking sequence, increasing the level of controllability and symmetry in these pores. Furthermore, we observe that pores with bilayer edges are preferentially formed, which is most likely not due to interlayer covalent bonds but rather kinetic effects. We observe a variety of monolayer and bilayer pores in bilayer AB-*h*-BN in our HRTEM focal series reconstructions and find excellent matches from theoretical simulations. These pores have a variety of electronic properties, ranging from half-metallic to semiconducting, which is encouraging for future research toward many applications such as DNA sequencing, molecular sieving and quantum emitters.

**Data Availability Statement**

The data that support the findings of this study are available from the corresponding author upon reasonable request.

**Supplementary Material**

See supplementary material for the details of the experimental and theoretical methods, 3 supplementary tables, and 5 supplementary figures.

**Acknowledgements**

This work was supported primarily by the Director, Office of Science, Office of Basic Energy Sciences, Materials Sciences and Engineering Division, of the U.S. Department of Energy under contract No. DE-AC02-05-CH11231, within the sp2-bonded Materials Program (KC2207), which supported TEM imaging and first-principles computations of the atomic structures. Sample growth was supported by the Director, Office of Science, Office of Basic Energy Sciences, Materials Sciences and Engineering Division, of the U.S. Department of Energy under contract No. DE-AC02-05-CH11231, within the Van der Waals Heterostructures Program (KCWF16). Further support for theoretical work was provided by the NSF Grant No. DMR-1926004 which supported first-principles computations of the precise electronic structures. Computational resources were provided by the DOE at Lawrence Berkeley National Laboratory's NERSC facility and the NSF through XSEDE resources at NICS. SMG acknowledges support from the Kavli Energy NanoSciences Institute Fellowship and the NSF Graduate Fellowship Program. MD thanks Sehoon Oh and PE thanks Earl J. Kirkland for helpful scientific discussions.

# Supplementary Material for
# Electron Beam-Induced Nanopores in Bernal-Stacked Hexagonal Boron Nitride

Mehmet Dogan[1,2] [(*)], S. Matt Gilbert[1,2,3] [(*)], Thang Pham[1,2,3,4], Brian Shevitski[1,3,5], Peter Ercius[5], Shaul Aloni[5], Alex Zettl[1,2,3,#], Marvin L. Cohen[1,2,#]

[1] Department of Physics, University of California, Berkeley, CA 94720, USA
[2] Materials Sciences Division, Lawrence Berkeley National Laboratory, Berkeley, CA 94720, USA
[3] Kavli Energy NanoScience Institute at the University of California, Berkeley, and the Lawrence Berkeley National Laboratory, Berkeley, CA 94720, USA
[4] Department of Materials Science and Engineering, University of California, Berkeley, CA 94720, USA
[5] Molecular Foundry, Lawrence Berkeley National Laboratory, Berkeley, CA 94720, USA
[(*)] These authors contributed equally to this work.
[(#)] To whom correspondence should be addressed: azettl@berkeley.edu, mlcohen@berkeley.edu

**Experimental Methods**

The growth and structural characterization procedures are described in detail in Reference 1. In short, we grow Bernal-stacked (AB-)*h*-BN on Cu and Fe substrates by low-pressure chemical vapor deposition (LP-CVD). We employ a two-zone heating approach, in which the gaseous thermal decomposition products of solid ammonia borane precursors (kept at 70-90 °C) react to form h-BN on a transition metal catalyst (heated at 1025 °C). We verify the Bernal-stacking of the resulting multilayer flakes using high-resolution transmission electron microscopy (HRTEM) and selected area electron diffraction (SAED). HRTEM was carried out at 80 kV using the TEAM 0.5 microscope at Lawrence Berkeley National Laboratory equipped with image aberration correctors. Exit wave reconstructions of the HRTEM focal series (**Figure 2** and **Figure 3** in the main text) are produced using Mac Tempas software (defocus values of -10 nm to 70 nm with a step of 1 nm).

The fabrication of the nanopores is described in detail in Reference 2. In short, the 80 kV electron beam is condensed to a 10–20 nm diameter at spot size 3, alpha = 3 with a beam current of 37 A/cm$^2$. One pore forms in a newly exposed layer roughly every 5 minutes. So, we monitor and observe the formation of new defects by frequently spreading the beam and acquiring images. Once the formation of a vacancy is observed, the beam is left expanded at a beam current of 3 A/cm$^2$ at spot size 3 or 6 A/cm$^2$ at spot size 1, which leads to the gradual enlargement of the pore. The typical behavior of the pore area versus the electron beam dose (current per area × time) is shown in the Figure 3 of Reference 2.

**Theoretical Methods**

We employ density functional theory (DFT) in the Perdew–Burke–Ernzerhof generalized gradient approximation (PBE GGA) to conduct the first-principles calculations.[3] We implement

DFT using the QUANTUM ESPRESSO software package with norm-conserving pseudopotentials.[4,5] We set 120 Ry as the plane-wave energy cutoff for the pseudo Kohn-Sham wavefunctions. For single- and double-layer nanopore simulations, we use a 12×12×1 Monkhorst-Pack k-point mesh per 1×1 unit cell (u.c.) to sample the Brillouin zone (3×3×1 k-point mesh for 5×5 u.c., 2×2×1 k-point mesh for 6×6 u.c., 7×7 u.c. and 8×8 u.c.). For edge simulations, we use a 12×1×1 k-point mesh. All atomic coordinates are relaxed until the forces on all the atoms are less than $10^{-3}$ Ry/$a_0$ in all three Cartesian directions ($a_0$: Bohr radius). A ~14 Å thick vacuum is used between the periodic copies of the slab in the out-of-plane direction. For edge simulations, a 1×10 cell is constructed, and ~12 Å of vacuum is placed between the copies of the 1D system along the direction perpendicular to the edge. We passivate the dangling bonds at the other edge, which is not of interest to us, using hydrogen atoms, and keep the first 4 unit cells (8 atoms) unrelaxed. In order to include the interlayer van der Waals (vdW) interactions, we include a Grimme-type dispersion correction.[6]

HRTEM simulations are performed using the autoslic program from E. J. Kirkland's Computem multislice programs.[7] The simulation settings are 80 kV acceleration voltage, 512×512 pixels, 40×40 Å supercell size. The final exit wave was convoled with a Gaussian function of 1.25 sigma (~23 picometer FWHM) to approximate lattice vibrations. The exit waves are then binned by two to compare with the experimental pixel size. The phase of the exit wave is shown in **Figure S3** middle row.

We use 5×5 unit cells to compute the boron monovacancy, and increase the unit cell size by one in both directions for each step in increasing the nanopore size. This keeps the distance between neighboring nanopores constant for all nanopore sizes. To ensure that our description of the atomic positions, electronic structure and magnetic properties are converged with respect to the unit cell size, we have conducted a convergence study for the single-layer boron monovacancy, which is presented in **Table S2, Table S3 and Figure S5**. We find that a 5×5 u.c. leads to sufficiently converged results, and although using a 6×6 u.c. would increase numerical accuracy, starting from a 5×5 u.c. and going up to an 8×8 u.c. allows us to study larger nanopores with reasonable computational cost.

**Table S1. Energies of the interlayer-bonded structures for boron monovacancies.** Energies of the interlayer-bonded structures, relative to the configuration in which out-of-plane relaxations are not allowed, for the boron monovacancy in one layer of a bilayer AA′-stacked and AB1-stacked *h*-BN. The calculations are conducted in electron-rich conditions. For each stacking sequence, formation of 1, 2 or 3 interlayer bonds (ILBs) is possible between the pairs of N atoms at the edge of the vacancy in the top layer and the B atoms directly underneath; so, each case corresponds to a (meta)stable configuration.

|  | $\Delta E$ (eV) | | |
| --- | --- | --- | --- |
|  | 1 ILB | 2 ILBs | 3 ILBs |
| AA′ | –1.09 | –1.49 | –1.08 |
| AB1 | –1.03 | –1.36 | –1.07 |

**Table S2. Convergence of the structure and total magnetization with respect to unit cell size.** For the single-layer boron monovacancy, each of the following quantities are listed for simulation cells varying from 5×5 to 8×8: $\Delta E_1$ is the total energy difference between the relaxed atomic configuration and the initial atomic configuration where all the atoms are fixed to the perfect hexagonal lattice positions (with a boron atom taken out of the structure). **Edge B-N-B angle** is angle between the two bonds th edge N atom makes with its B neighbors. For comparison, this angle is 126.4° for the single-layer N-terminated edge. $\Delta E_2$ is the reduction in energy caused by allowing magnetization in the calculation. $M_{\text{total}}$ is the total magnetization of the unit cell. For comparison, the total magnetization of the single-layer N-terminated edge is 1.01 $\mu_B$.

| Simulation cell | $\Delta E_1$ (eV) | Edge B-N-B angle | $\Delta E_2$ (eV) | $M_{\text{total}}$ ($\mu_B$) |
|---|---|---|---|---|
| 5×5 | –0.491 | 129.2° | –0.316 | 3.23 |
| 6×6 | –0.483 | 129.2° | –0.319 | 3.32 |
| 7×7 | –0.486 | 129.3° | –0.322 | 3.34 |
| 8×8 | –0.489 | 129.3° | –0.324 | 3.35 |

**Table S3. Convergence of the local magnetization with respect to unit cell size.** For the single-layer boron monovacancy, magnetizations around the atoms near the nanopore are listed for simulation cells varying from 5×5 to 8×8. The labeling of the atoms is shown in **Figure S5(a)**. For the single-layer N-terminated edge, $M_{\text{N(\#1)}} = 0.93\ \mu_B$, $M_{\text{B(\#2)}} = -0.10\ \mu_B$ and $M_{\text{N(\#3)}} = 0.16\ \mu_B$ (see **Figure S5(b)** for the labeling of atoms). The magnetization is computed by decomposing the Löwdin charges on the atomic orbitals corresponding to each atom.

| Simulation cell | $M_{\text{N(\#1)}}$ ($\mu_B$) | $M_{\text{B(\#2)}}$ ($\mu_B$) | $M_{\text{N(\#3)}}$ ($\mu_B$) | $M_{\text{N(\#4)}}$ ($\mu_B$) |
|---|---|---|---|---|
| 5×5 | 0.84 | –0.04 | 0.07 | 0.09 |
| 6×6 | 0.86 | –0.04 | 0.07 | 0.10 |
| 7×7 | 0.87 | –0.04 | 0.07 | 0.10 |
| 8×8 | 0.87 | –0.04 | 0.07 | 0.10 |

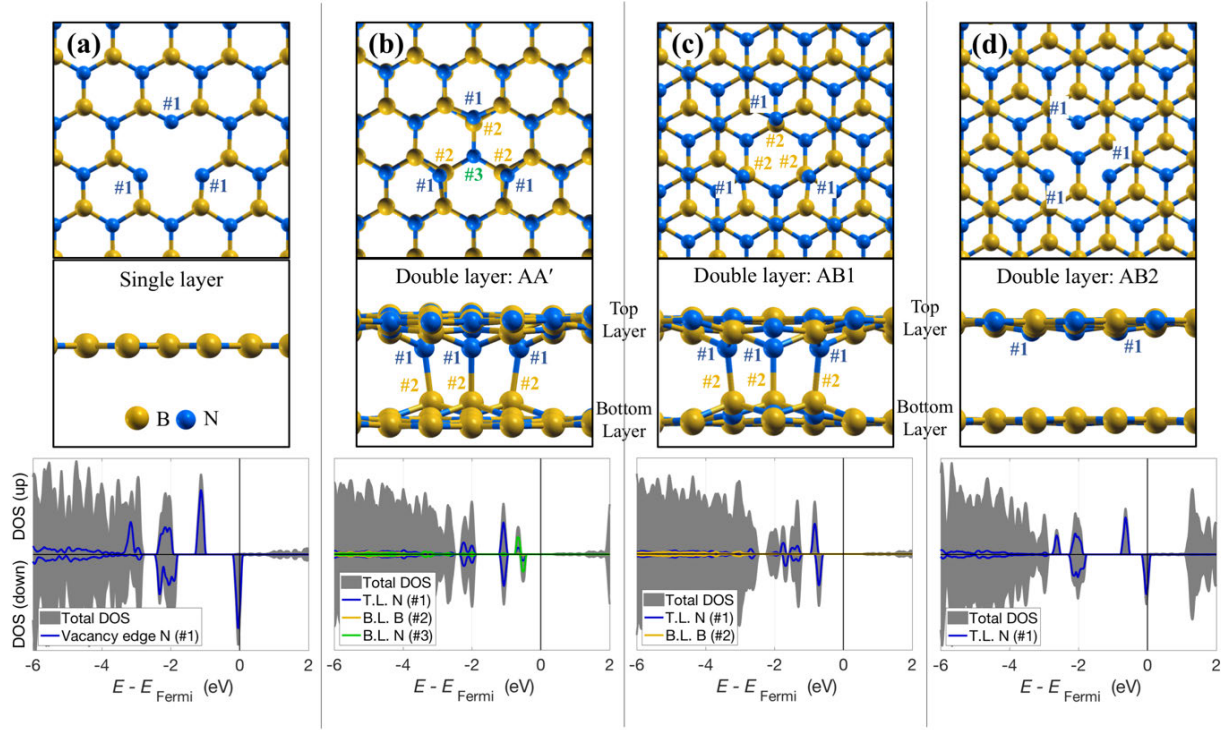

**Figure S1. Simulated boron monovacancies in monolayer and bilayer h-BN.** Relaxed atomic structure and density of states (DOS) of a boron monovacancy in (a) monolayer, (b) AA′-stacked bilayer, (c) AB1-stacked bilayer and (d) AB2-stacked bilayer h-BN. For each structure, top and side views of the atomic configuration are presented at the top. The calculations are carried out in electron-rich conditions (3 extra electrons per 5×5 cell). Boron is shown in gold and nitrogen is shown in blue. Computed spin-resolved density of states (DOS) plots for each structure are presented at the bottom. The total DOS is plotted as gray areas, and the colored curves are obtained by adding up the DOS projected onto the 2s and 2p orbitals (PDOS) of the atoms marked on the figure: (a) 3 edge N atoms (blue); (b) 3 edge N atoms in the top layer (blue), 3 B atoms directly underneath (gold), 1 N atom in the bottom layer directly under the vacancy (green); (c) 3 edge N atoms in the top layer (blue), 3 B atoms directly underneath (gold); (d) 3 edge N atoms in the top layer (blue).

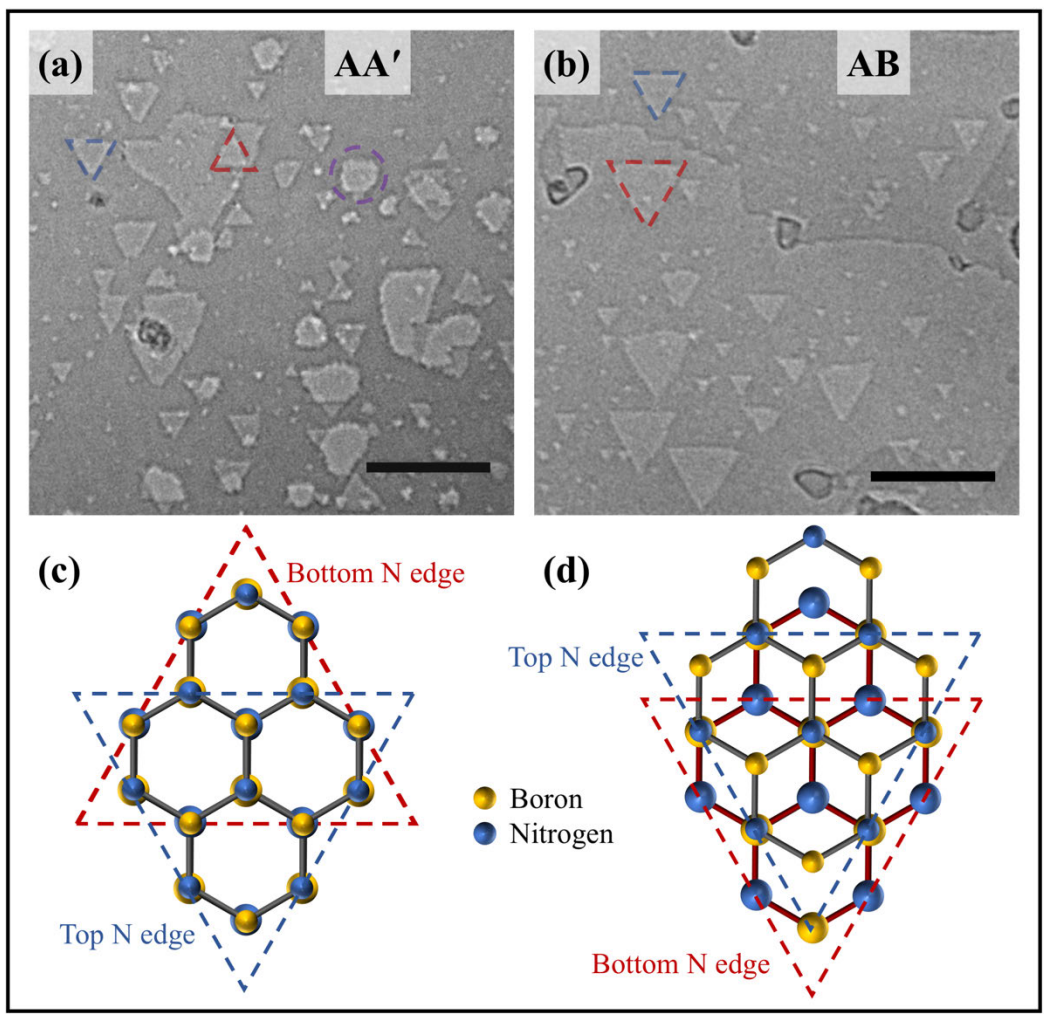

**Figure S2. Vacancies in AA′- and AB-stacked h-BN.** (a) and (b) show conventional TEM images of vacancies formed under 80 kV electron irradiation in AA′- and AB-stacked h-BN respectively. (a) In the AA′-stacked h-BN, we observe anti-parallel triangular vacancies in separate layers, highlighted by red and blue triangles. We also observe bilayer vacancies with no preferred shape or edge termination (purple circle). (b) In the AB-stacked h-BN, we observe only parallel triangles in every layer as highlighted by red and blue triangles. (c) A schematic of bilayer AA′-stacked h-BN. The nitrogen zig-zag edges are highlighted in the top (blue) and bottom (red) layers. (d) A schematic of bilayer AB-stacked h-BN. The nitrogen zig-zag edges are highlighted in the top (blue) and bottom (red) layers. Boron is shown in gold and nitrogen is shown in blue.

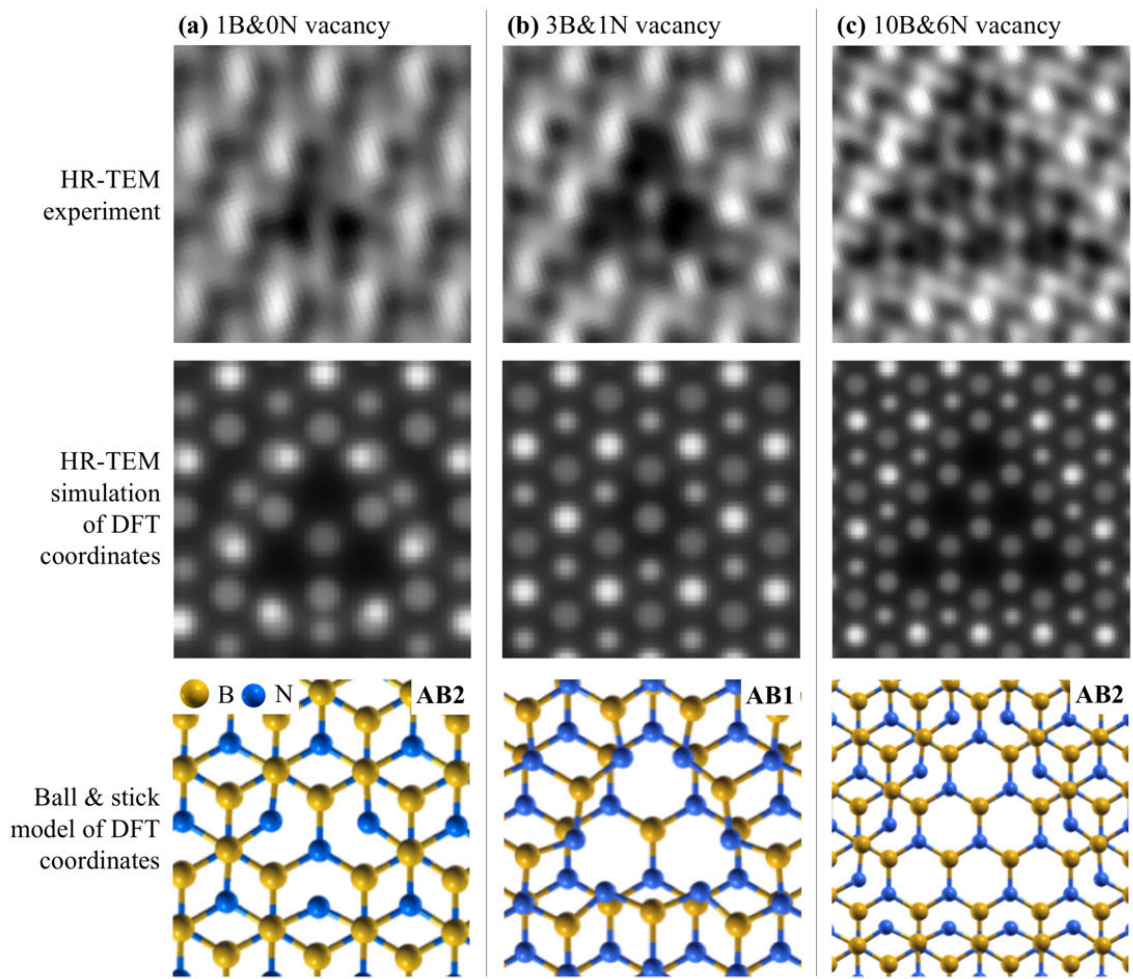

**Figure S3. Vacancies in one layer of a bilayer Bernal-stacked h-BN.** Top row: HRTEM focal series reconstructions of vacancies produced in a single layer of a bilayer AB-stacked h-BN, comprised of (a) 1 missing boron atom (monovacancy), (b) 3 missing boron atoms and 1 missing nitrogen atom (tetravacancy), and (c) 10 missing boron atoms and 6 missing nitrogen atoms. Middle row: HRTEM simulations (the phase of the exit wave) of the same systems using the optimized atomic coordinates obtained from DFT calculations. Bottom row: Ball & stick models of the computed atomic structures of the same systems. Boron is shown in gold and nitrogen is shown in blue.

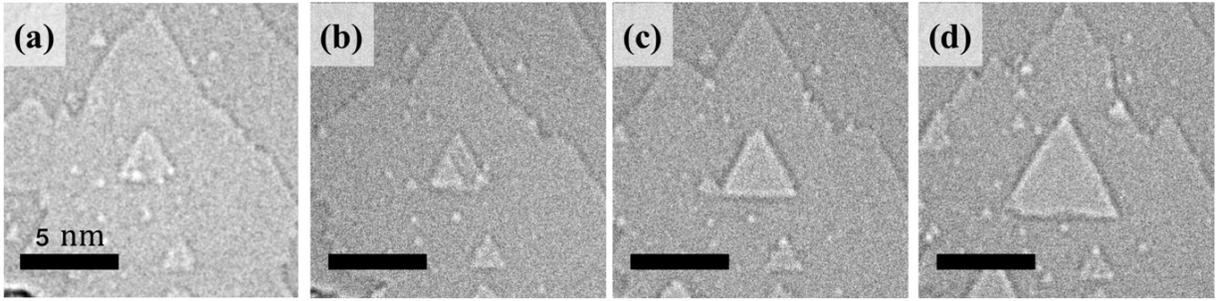

**Figure S4. Growth of a bilayer nanopore in AB-stacked h-BN.** (a)-(d) shows the growth of a bilayer nanopore from a few-atom monolayer vacancy to a 6 nm bilayer pore. (a) Initially, the vacancy starts as a one-edge bilayer few-atom defect. (b) Then, the pore grows to ~1.5 nm with two bilayer edges. (c) The vacancy grows to 4 nm with three bilayer edges. (d) The pore grows to 6 nm retaining its bilayer edges. No edge is ever observed to revert to monolayer. All scale bars are 5 nm. Images are taken at roughly 2-minute intervals.

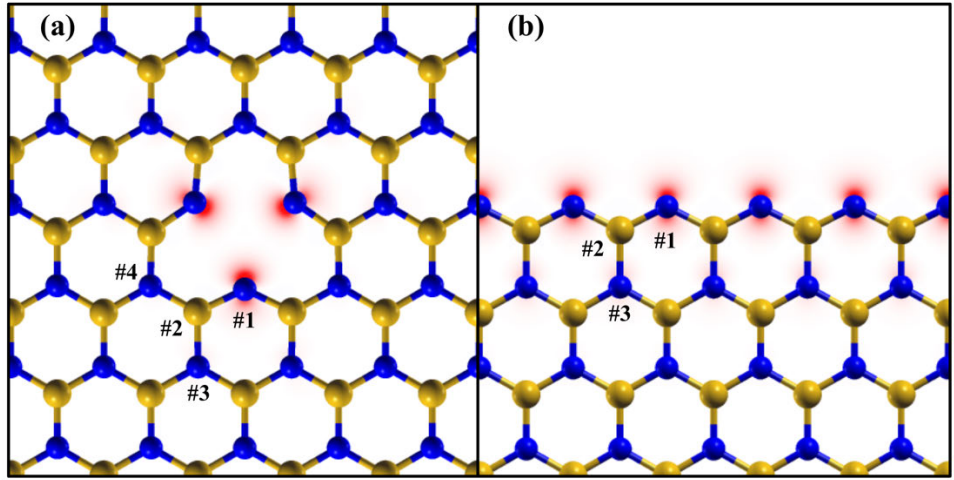

**Figure S5. Localization of the magnetization for a vacancy and an edge.** (a) and (b) show the spatial distribution of the magnetization for a boron monovacancy and an N-terminated edge, respectively, in a single sheet of *h*-BN. Boron and nitrogen atoms are displayed in gold and blue color, respectively. **Table S3** lists the magnetization values computed by decomposing the Löwdin charges on the atomic orbitals corresponding to each atom labeled in this figure.